\documentclass{JHEP3}
\usepackage[dvips]{graphicx}  

\def\be{\begin{eqnarray}}
\def\ee{\end{eqnarray}}
\def\bea{\begin{eqnarray}}
\def\eea{\end{eqnarray}}

\newcommand{\eq}[1]{Eq.~(\ref{eq:#1})}
\newcommand{\sect}[1]{Sec.~\ref{sec:#1}}
\newcommand{\appen}[1]{App.~\ref{sec:#1}}
\newcommand{\fig}[1]{Fig.~\ref{fig:#1}}
\newcommand{\tabl}[1]{Table~\ref{table:#1}}

\newcommand{\she}{\sinh^2{\eta}}
\newcommand{\che}{\cosh^2{\eta}}

\newcommand{\calF}{{\cal F}}

\newcommand{\largeeta}{\xrightarrow{\eta \gg 1}}

\newcommand{\sx}{\sigma_x}
\newcommand{\sr}{\sigma_r}
\newcommand{\sgh}{\sigma_h}
\newcommand{\deltakt}{\delta_{\rm cascade}}
\newcommand{\deltadp}{\delta_{Dp}}

\newcommand{\SAdS}{${\rm SAdS}_5\:$}
\newcommand{\SAdSd}{${\rm SAdS}_{d+1}\:$}

\usepackage{amsmath,amsfonts,amssymb}
\numberwithin{equation}{section}
\newcommand{\calE}{\mathcal{E}}
\newcommand{\calI}{\mathcal{I}}
\newcommand{\calJ}{K}
\newcommand{\calW}{\mathcal{W}}

\newcommand{\hp}{\check{p}}
\newcommand{\hq}{\check{q}}
\newcommand{\tilp}{\Tilde{p}}
\newcommand{\tilq}{\Tilde{q}}
\newcommand{\tilI}{\Tilde{I}}
\newcommand{\tiltheta}{\Tilde{\theta}}

\title{Screening length and the direction of plasma winds}

 \author{Makoto Natsuume\\Theory Division, Institute of Particle 
and Nuclear Studies \\
KEK, High Energy Accelerator Research
Organization\\ Tsukuba, Ibaraki, 305-0801, Japan\\
\email{makoto.natsuume@kek.jp}}

\author{Takashi Okamura\\
Department of Physics\\   Kwansei Gakuin University\\
Sanda, Hyogo, 669-1337, Japan\\
\email{okamura@ksc.kwansei.ac.jp}}

\preprint{KEK-TH-1155} 

\abstract{
We study the screening length of a heavy quark-antiquark pair in strongly coupled gauge theory plasmas flowing at velocity $v$ following a proposal by Liu, Rajagopal, and Wiedemann. 
We analyze the screening length as the direction of the plasma winds vary. To leading order in $v$, this angle-dependence can be studied analytically for many theories by extending our previous formalism. We show that the screening length is locally a minimum (maximum) when the pair is perpendicular (parallel) to the plasma winds, which has been observed for the ${\cal N}=4$ plasma.
Also, we compare AdS/CFT results with weak coupling ones, and we discuss the subleading dependence on $v$ for the D$p$-brane. 
}

\keywords{AdS/CFT correspondence, Brane dynamics in gauge theories}

\begin{document}

\section{Introduction}

For the last couple of years, many authors try to apply the AdS/CFT correspondence \cite{Maldacena:1997re,Gubser:1998bc,Witten:1998qj,Witten:1998zw} to the real quark-gluon plasma (QGP) system. (See Ref.~\cite{Natsuume:2007qq} for a review.) 
Recently, there has been an  interesting proposal by Liu, Rajagopal and
Wiedemann to model $J/\Psi$-suppression in the QGP medium via the  correspondence \cite{Liu:2006nn}.%
\footnote{Refs.~\cite{Chernicoff:2006hi,Peeters:2006iu} made independent proposals. See also Refs.~\cite{Caceres:2006ta}-\cite{Liu:2006he} for extensions of the proposal.}
To incorporate the effect of moving quarks relative to the plasma, the authors  considered  a boosted black hole and computed
the screening length in the quark-antiquark rest frame. 
The main lessons drawn from Ref.~\cite{Liu:2006nn} are,
\begin{enumerate}
\item[(i)] The screening length is proportional to $(\mbox{boosted energy density})^{-1/4}$.
\item[(ii)] Aside from the boost factor, $(1-v^2)^{1/4}$, \ the screening length has only a mild dependence on the plasma wind velocity $v$.
\item[(iii)] The screening length has the minimum when the quark-antiquark pair (dipole) is perpendicular to the plasma wind ($\theta=\pi/2$) and has the maximum when the dipole is parallel to the wind ($\theta=0$). 
\end{enumerate}
The main focus in Ref.~\cite{Liu:2006nn} is the ${\cal N}=4$ super-Yang-Mills theory (SYM) at finite temperature or Schwarzschild-${\rm AdS}_5$ black holes (${\rm SAdS}_5$). Even for such a simple theory,  numerical computations were needed in order to see the full details of the screening length.

In our previous paper~\cite{Caceres:2006ta}, we have studied property~(i) above. We focus on the ultrarelativistic limit, where analytic
computations are possible.
This makes it easier to carry out the analysis in
various, more involved, backgrounds. This is important since  it is not clear which properties of the screening length found in Ref.~\cite{Liu:2006nn} are generic given only the \SAdS example. 
In the ultrarelativistic limit, 
\be
(\mbox{screening length}) \propto (1-v^2)^{\nu}~,
\ee
and we have computed the exponent $\nu$ in various theories.

One may criticize that this limit is not a realistic situation for heavy quarks, but this is not our point. Since the screening length has only a mild dependence on $v$ aside from the boost factor [property~(ii)], this limit serves a good approximation to a more realistic situation. (See \sect{subleading})

In this paper, we study property~(iii), which was not covered in our previous paper. We study the screening length as the direction of the plasma winds vary. This $\theta$-dependence has been studied for \SAdS in Ref.~\cite{Liu:2006nn} numerically, but we discuss it analytically in the ultrarelativistic limit. It turns out that many theories behave similarly to the ${\rm SAdS}_5$ case. Namely, 
\begin{enumerate}
\item We study the $\theta$-dependence of the screening length near $\theta = \pi/2$ (\sect{perpendicular}). We show that the screening length at $\theta=\pi/2$ is locally a minimum generically. This is the case for all theories which satisfy the conditions below~(\ref{eq:fall-off}). In particular, this is true for R-charged black holes \cite{Kraus:1998hv,Cvetic:1999xp,Behrndt:1998jd}, the ${\rm SAdS}_{d+1}$ black holes, the nonextreme Klebanov-Tseytlin geometry \cite{Gubser:2001ri,Buchel:2001gw,Buchel:2000ch}, and the D$p$-brane. 
\item Similarly, we study the $\theta$-dependence of the screening length near $\theta = 0$ (\sect{parallel}). We show that the screening length at $\theta=0$ is locally a maximum generically. 
\end{enumerate}

%
%
In the next section, we briefly summarize our procedure to derive the screening length developed in Ref.~\cite{Caceres:2006ta}. In \sect{discussion}, we comment on three other issues. First, we compare the AdS/CFT results with the results by Chu and Matsui, who studied this problem in a Abelian gauge theory  \cite{Chu:1988wh}. Second, we study property~(ii) for the D$p$-brane. Finally, we discuss the relation between $\nu$ and the speed of sound.

\section{Setup}\label{sec:setup}


\begin{figure}[tb]
\begin{center}
\includegraphics{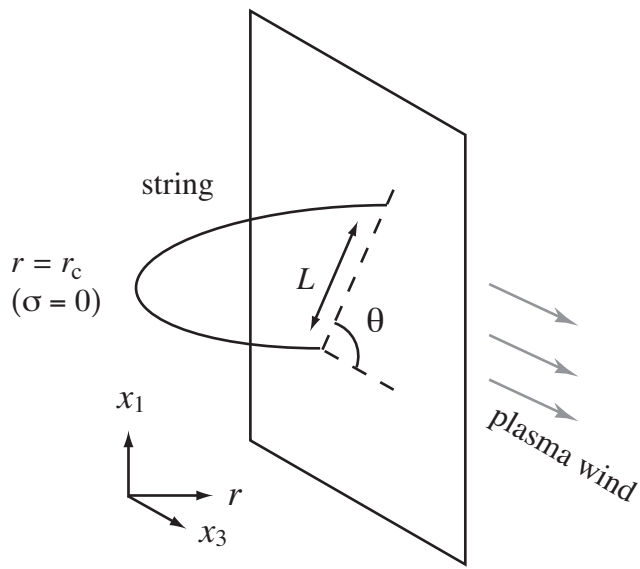}
\vskip2mm
\caption{The fundamental string connecting the quark-antiquark pair. (The shape of the string should not be taken seriously.) $L$ is the dipole length.}
\label{fig:wind}
\end{center}
\end{figure}


In order to consider the screening length in the dipole rest frame, we boost the  background metric. We assume an unboosted metric of the form
\be
ds^2 = g_{xx}\,\{-(1-h)dt^2+dx_i^2\}+g_{rr}\,dr^2 +\cdots~.
\ee
The quark-antiquark pair is chosen to lie in the $(x_1,x_3)$-plane at an angle $\theta$ relative to the wind, 
which we choose the $x_3$-direction (See \fig{wind}). 
Thus, we choose the gauge %
\footnote{This fails to be a good gauge when $\theta=0$, and one needs to interchange $x_1$ and $x_3$, but the final formulae remain valid even for $\theta=0$.}
$\tau = t,~\sigma = x_1$ and consider the configuration
\be
  x_3 = x_3(\sigma)~, \quad 
  r = r(\sigma)~, \quad 
  \mbox{constant otherwise}.
\label{eq:configuration}
\ee
The world-sheet configuration can be obtained by solving the equations of motion from the Nambu-Goto action (in the string metric). The solution is parametrized by two integration constants, say, $p$ and $q$:
\bea
q^2 r'^2 &=& \frac{g_{xx}}{g_{rr}} \left[
\frac{1-h \che}{1-h}\left\{ 
g_{xx}^2(1-h)-p^2 \right\} - q^2 \right]
=: \calF (r)~,
\label{eq:EOM_r} \\
q^2 x_3'^2 &=& p^2\,\left(\frac{1-h \che}{1-h}\right)^2~,
\label{eq:EOM_x3}
\eea
Here, $~{}' = d/d\sigma$, $\cosh\eta = \gamma$ and $\gamma=1/\sqrt{1-v^2}$, where $v$ is the wind velocity. Note that the equations of motion possess the symmetry $\sigma\rightarrow-\sigma$.
As the boundary conditions, we consider the string which stretches from  asymptotic infinity and reaches a turning point
$r=r_c$ defined by $\calF(r_c)=0$. 
Then, the string goes back to  asymptotic
infinity. From the symmetry  $\sigma\rightarrow-\sigma$, the turning point occurs at $\sigma=0$.  
These boundary conditions determine the
integration constants $p$ and $q$ in terms
of $L$ and $\theta$:
\bea
  \frac{L}{2}\, \sin\theta
  &=& q\, I_s(p, q, \eta)~,
\label{eq:L_sin} \\
  \frac{L}{2}\, \cos\theta
  &=& p\, \Big[~I_s(p, q, \eta)
  - \sinh^2\eta\, I_c(p, q, \eta)~\Big]~,
\label{eq:L_cos}
\eea
where
\bea
  I_s(p, q, \eta)
  &:=& \int^\infty_{r_c} \frac{dr}{ \sqrt{ \calF } }~,
\label{eq:def-I_s} \\
  I_c(p, q, \eta)
  &:=& \int^\infty_{r_c} \frac{dr}{ \sqrt{ \calF } }\,
                       \frac{h}{ 1 - h }~.
\label{eq:def-I_c}
\eea
We are interested in the behavior of $L$ as a function of $(p, q)$ for a fixed $\theta$. For simplicity, consider the $\theta=\pi/2$ case which corresponds to $p=0$. The length $L$ goes to zero both for small $q$ and for large $q$. Thus, there is a maximum value $L_s$ at some $q_m$. This means that there is no extremal world-sheet which binds the quark-antiquark pair for $L>L_s$, so this $L_s$ is defined as a screening length in Ref.~\cite{Liu:2006nn}.

The energy is given by
\bea
E &=& -\frac{S}{\cal T}  \nonumber \\
&=& \frac{1}{\pi l_s^2} \int_{r_c}^\infty dr\, 
\sqrt{g_{xx}g_{rr}(1-h \che)}
\sqrt{ \frac{g_{xx}}{g_{rr}\calF} (q^2+p^2\frac{1-h \che}{1-h})+1 }~,
\label{eq:energy}
\eea
where $S$ is the action for the dipole, and ${\cal T}$ is the proper time extension of the dipole. 
As usual, this energy can be made finite by subtracting the self-energy of 
a disconnected quark and antiquark pair (See \sect{chu-matsui}).

We assume that 
the metric falls off as
\begin{align}
  & g_{xx}(r)
  \sim \left( \frac{r}{R} \right)^{\sigma_x}~,
& & g_{rr}
  \sim C^2\, \left( \frac{r}{R} \right)^{-\sigma_r}~,
& & h(r) \sim \frac{m}{r^{\sigma_h}}
  = \frac{m}{R^{\sigma_h}}\, \left( \frac{r}{R} \right)^{-\sigma_h}~,
\label{eq:fall-off}
\end{align}
near the infinity $r = \infty$, where $R$ is the AdS radius.
The parameter ``$m$" is the mass parameter which represents the energy density of the (unboosted) black holes in our examples.
Furthermore, we assume that the metric behaves as
\begin{align}
  & g_{xx}^2~h \sim \left(\text{at most~~} O(1)\right)~,
& & g_{xx} \sim (\text{divergent})~,
\label{eq:fall-off_index}
\end{align}

If $\sigma_x,~\sigma_h > 0$,
the turning point satisfies
$h(r_c) \ll 1$ for large-$\eta$ 
so that the turning point is near the infinity.
We are interested in the leading order term of $\cosh\eta$,
so we need only the leading term of the metric.

Using the parameter,
\begin{align}
  & \calE := \frac{m \cosh^2\eta}{R^{\sigma_h}}~,
\label{eq:def-calE}
\end{align}
and the rescaled variables
\begin{align}
  & t := \frac{ (r/R)^{\sigma_h} }{\calE}~,
& & \tilp^2 := \frac{p^2}{\calE^{2 \sigma_x/\sigma_h}}~,
& & \tilq^2 := \frac{q^2}{\calE^{2 \sigma_x/\sigma_h}}~,
\label{eq:def-new_variables}
\end{align}
we can rewrite Eqs.~(\ref{eq:L_sin}) and (\ref{eq:L_cos}) as
\begin{align}
  & \frac{L}{R}\, \sin\theta
  \largeeta \frac{2\, C}{\sigma_h}~\calE^{-\nu}~\tilq~
  \tilI_s(\tilp, \tilq)~,
\label{eq:L_sin-approx} \\
  & \frac{L}{R}\, \cos\theta
  \largeeta \frac{2\, C}{\sigma_h}~\calE^{-\nu}~\tilp~
     \Big[~\tilI_s(\tilp, \tilq) - \tilI_c(\tilp, \tilq)~\Big]~,
\label{eq:L_cos-approx}
\end{align}
where
\begin{align}
  & \tilI_s(\tilp, \tilq)
  := \int^\infty_{t_c} dt~\frac{ t^{-\nu-1/2} }
    { \sqrt{ ( t - 1 ) \left( t^{2 \lambda}
            - \tilp^2 \right)- \tilq^2\, t } }~,
\label{eq:def-tilI_s} \\
  & \tilI_c(\tilp, \tilq)
  := \int^\infty_{t_c} \frac{dt}{t}~
    \frac{ t^{-\nu-1/2} }
         { \sqrt{ ( t - 1 ) \left( t^{2 \lambda}
                - \tilp^2 \right)- \tilq^2\, t } }~,
\label{eq:def-tilI_c} 
\end{align}
and
\begin{align}
  & \nu := \frac{\sx+\sr-2}{2\sgh}~,
& & 0 < \lambda := \frac{\sigma_x}{\sigma_h} \le \frac{1}{2}~.
\label{eq:def-index}
\end{align}
The turning point $t_c \ge \text{max}(|\tilp|^{1/\lambda}, 1)$
is then determined by
\begin{align}
  & 0 =  ( t_c - 1 ) \left( t_c^{2 \lambda}
      - \tilp^2 \right)- \tilq^2\, t_c~.
\label{eq:turning_pt0}
\end{align}
%

Equations~(\ref{eq:L_sin-approx}) and (\ref{eq:L_cos-approx}) imply that 
the maximum of $L$, $L_s$, behaves as
\begin{align}
  & L_s
  \propto R \calE^{-\nu}
  \propto R \left( \frac{m}{R^{\sigma_h}}~\cosh^2\eta \right)^{-\nu}
\label{eq:L_max-behavior}
\end{align}
irrespective of $\theta$. 
Since the parameter $m$ is related to the (unboosted) energy density in our examples, the screening length $L_s$ is written in terms of
the boosted ``energy density" of plasma wind 
at large-$\eta$.

Our parametrization makes use of the energy density, but one can adopt the temperature as in Ref.~\cite{Liu:2006nn}. In this case, the screening length is inversely proportional to temperature for ${\rm SAdS}_{d+1}$ and the D$p$-brane. 

On the other hand, for R-charged black holes, it is very natural to write the screening length in terms of energy density \cite{Caceres:2006ta}. In terms of temperature, the screening length would be very complicated. The parametrization in terms of the energy density is favored in this case.

\tabl{exponent} summarizes the exponent $\nu$ for various theories obtained in Ref.~\cite{Caceres:2006ta}.

\begin{table}
\begin{center}
\begin{tabular}{|l||c|c|c||c|}
\hline
Backgrounds	& $\sx$	& $\sr$ & $\sgh$ & $\nu$ \\
\hline
&&&& \\
D3-brane  & 2 & 2 & 4 & $\displaystyle{\frac{1}{4}}$ \\
&&&& \\
R-charged & 2 & 2 & 4 & $\displaystyle{\frac{1}{4}}$ \\
&&&& \\
$\text{SAdS}_{d+1}$ & 2 & 2 & $d$ & $\displaystyle{\frac{1}{d}}$ \\
&&&& \\
D$p$-brane & $\displaystyle{\frac{7-p}{2}}$ 
     & $\displaystyle{\frac{7-p}{2}}$ 
     & $7-p$ 
     & $\displaystyle{\frac{5-p}{2(7-p)}}$ \\
&&&& \\
Klebanov-Tseytlin & $2-\deltakt$ 
   & $2-\deltakt$ 
   & $4$ 
   & $\displaystyle{\frac{1-\deltakt}{4}}$ \\
&&&& \\
\hline
\end{tabular}
\caption{The fall-off behavior of the metric and the scaling exponent $\nu$ for various theories. For the KT geometry, the $\sigma$'s are the values in the standard Schwarzshild-like radial coordinates $r$, (The $\sigma$'s depend on the choice of radial coordinates, but the exponent $\nu$ does not depend on the choice. See Ref~\cite{Caceres:2006ta} for the details) and $\deltakt$ is the deformation parameter from the conformal theory. }
\label{table:exponent}
\end{center}
\end{table}

\section{$\theta$-dependence}\label{sec:theta}

\subsection{Preliminaries}

We estimate the $\theta$-dependence of the screening length.
One obstacle is that the integration constants $\tilp$ and $\tilq$
are determined by the dipole length $L$ and $\theta$, 
but their relations are not transparent.
We further choose the variables
so that the relations become more transparent.

We change the integration variable
from $t$ to $u:=t/t_c$ in Eqs.~(\ref{eq:L_sin-approx})
and (\ref{eq:L_cos-approx}):
\begin{align}
  & \frac{L}{R}\, \sin\theta
  \sim A~\hq~t_c^{-\nu}~\calI_s(\hp, \hq)~,
\label{eq:L_sin-approxII} \\
  & \frac{L}{R}\, \cos\theta
  \sim A~\hp~t_c^{-\nu}~
  \left( \calI_s(\hp, \hq) - \frac{\calI_c(\hp, \hq)}{t_c} \right)~,
\label{eq:L_cos-approxII}
\end{align}
where
\bea
  A &:=& \frac{2\, C}{\sigma_h}~\calE^{-\nu}~,
\label{eq:def-A} \\
  \hp &:=& \frac{\tilp}{t_c^{\lambda}}~,
\\
  \hq &:=& \frac{\tilq}{t_c^{\lambda}}~.
\label{eq:def-hp_hq}
\eea
The functions $\calI_s$ and $\calI_c$ are defined
in Eqs.~(\ref{eq:def-calI_s}) and (\ref{eq:def-calI_c})
in \appen{formulae}.
Below we use that
\be
\calI_c(\hp, \hq) < \calI_s(\hp, \hq)~.
\label{eq:Ic_vs_Is}
\ee

Let us look at the range of $\hp$ and $\hq$.
Recall that the turning point is defined
as $t_c \ge \text{max}(|\tilp|^{1/\lambda}, 1)$.
Then, $t_c^{2\lambda} \geq |\tilp|^2$ and $t_c \geq 1$.
The first relation is written as $\hp^2 \leq 1$.
Also, \eq{turning_pt0} can be written as
\be
\hq^2 =  ( 1 - 1/t_c )( 1 - \hp^2 )
\label{eq:turning_pt}
\ee
Using the above relations, one concludes $\hq < 1$.
Putting together these relations, one obtains $0 \leq \hp^2 \leq 1$ and $0 \leq \hq^2 < 1$.

\subsection{Behavior near $\theta = \pi/2$}\label{sec:perpendicular}

Here, we focus on the screening length $L_s$ near $\theta = \pi/2$.
We show that the screening length at $\theta=\pi/2$
is locally a minimum generically.
This is the case for all theories which satisfy
the fall-off conditions~(\ref{eq:fall-off}).


There are three constants $\hp, \hq$, and $t_c$. 
The screening length $L_s$ is given
by the maximum value of the dipole length $L$
as a function of these constants 
[under the constraints \eq{turning_pt} and $\theta=(\mbox{fixed})$].
We first eliminate $\hq$; 
then eliminate the constant $\hp$ using the angle $\theta$.
Finally, we vary $L$ as a function of $t_c^{-1}$
to obtain the maximum value of $L$.

First, we eliminate $\hq$ from
Eqs.~(\ref{eq:L_sin-approxII}) and (\ref{eq:L_cos-approxII})
using Eq.~(\ref{eq:turning_pt}):
\begin{align}
  & \frac{L}{R\, A}\, \sin\theta
  \simeq w^\nu~\sqrt{(1 - w)\, (1 - \hp^2)}~J_s(\hp^2, w)~,
\label{eq:L_sin-approxIII} \\
  & \frac{L}{R\, A}\, \cos\theta
  \simeq w^\nu~\hp~\Big[~J_s(\hp^2, w) - w~J_c(\hp^2, w)~\Big]~,
\label{eq:L_cos-approxIII}
\end{align}
where $0 < w := 1/t_c \le 1$.
The functions $J_s$ and $J_c$ are just $\calI_s$ and $\calI_c$, respectively, in terms of the variables $(\hp^2,w)$.
%
%
%
%
Introducing $\tiltheta := \pi/2 - \theta$,
these equations 
can be written as
\begin{align}
  & \frac{L}{R\, A}
  \simeq \frac{ \sqrt{(1 - w) (1 - \hp^2)} }{\cos\tiltheta}~w^\nu~
     J_s(\hp^2, w)~,
\label{eq:approx-Js-pi_half} \\
  & \tan\tiltheta
  \simeq \frac{\hp}{ \sqrt{(1 - w) (1 - \hp^2)} }~
   \left( 1 - w~\frac{ J_c(\hp^2, w) }{ J_s(\hp^2, w) } \right)~.
\label{eq:approx-tan_inv-pi_half}
\end{align}

Equation~(\ref{eq:approx-tan_inv-pi_half}) relates $\hp$ with the dipole angle $\theta$, so we eliminate $\hp$ by $\tiltheta$. Since $\hp\rightarrow 0$ as $\tiltheta\rightarrow 0$,%
\footnote{$1-w J_c/J_s \neq 0$ since $w \leq 1$ and $J_c/J_s<1$ from \eq{Ic_vs_Is}.}
\begin{align}
  & \hp = P(w)~\tiltheta + O(\tiltheta^3)~.
\label{eq:hp-approx-pi_half}
\end{align}
[See \eq{hp-approx-pi_half_app} for the definition of $P$.]
Then, $L$ is schematically written as
\be
\frac{L}{R A}
  = F_0(w) + F_1(w)\, \tiltheta^2 + O(\tiltheta^4)~,
\label{eq:L-approx-pi_half}
\ee
[See Eqs.~(\ref{eq:def-F_zero}) and (\ref{eq:def-F_one})
for the definitions of $F_0$ and $F_1$.]
Now, $L$ depends only on $w$ and on $\theta$
(which is fixed).
We evaluate the maximum value of $L$ as a function of $w$.

The value of $w$ which gives the maximum value of $L$
is denoted by $w_*$. Then, $w_*$ should satisfy
\begin{align}
  & 0
  = F'_0(w_*) + F'_1(w_*)\, \tiltheta^2 + O(\tiltheta^4)~.
\label{eq:eqn-w_star-pi_half}
\end{align}
As we vary the dipole angle $\theta$, $w_*$ changes.
Thus, one can expand $w_*$ as a power series in $\tiltheta$: 
$w_* =: w_0 + w_1\tiltheta^2 + O(\tiltheta^4)$.%
\footnote{
This power series expansion is implied
by the form of \eq{L-approx-pi_half} for $L$.}
Substituting this into \eq{eqn-w_star-pi_half}
gives the equations for $w_0$ and $w_1$:
\bea
  && F'_0(w_0) = 0~,
\label{eq:sol-w_star-pi_half0} \\
  && w_1 = - \frac{ F'_1(w_0) }{ F''_0(w_0)}~.
\label{eq:sol-w_star-pi_half1}
\eea
Equation~(\ref{eq:sol-w_star-pi_half0}) can be solved 
to obtain the solution $w_0$.
Then, $L_{\text{max}}$ is given by
\bea
   \frac{ L_{\text{max}} }{R A}
  &=& F_0(w_*) + F_1(w_*)\, \tiltheta^2 + O(\tiltheta^4)
\nonumber \\
  &=& F_0(w_0) + F'_0(w_0) w_1 \tiltheta^2
  + F_1(w_0)\, \tiltheta^2 + O(\tiltheta^4)
\nonumber \\
  &=& F_0(w_0) + F_1(w_0)\, \tiltheta^2 + O(\tiltheta^4)~.
\label{eq:L-approx-pi_halfII}
\eea

Thus, what is important to us is the sign of $F_1(w_0)$.
If this is positive, then the screening length
is locally a minimum at $\theta=\pi/2$.
%
The functions $F_0$ and $P^2$ are positive-definite
[The function $F_0$ is positive from the definition
of $L$ in \eq{L-approx-pi_half}],
so consider the sign of the combination $F_1/(F_0\, P^2)$.
Then, one can show
\begin{align}
   \frac{F_1(w)}{F_0(w)\, P^2(w)}
  &= \frac{w}{2 (1 - w)}~
      \left( 1 - \frac{J_c(0, w)}{J_s(0, w)} \right)^2
\nonumber \\
  &+ \frac{w}{2\, J_s(0, w)}~
      \left( \int^\infty_{1} \frac{ du~u^{-\nu-1/2} }
                { \sqrt{u - 1}~W^{3/2}(u, 0, w) }
  - \frac{J_c^2(0, w)}{J_s(0, w)} \right)~.
\label{eq:F_one}
\end{align}
Because the first term on the right-hand side is positive,
\begin{align}
   \frac{F_1(w)}{F_0(w)\, P^2(w)}
  &> \frac{w}{2\, J_s(0, w)}~
        \left( \int^\infty_{1} \frac{ du~u^{-\nu-1/2} }
                { \sqrt{u - 1}~W^{3/2}(u, 0, w) }
             - J_c(0, w) \right)~.
\label{eq:F_one-positivity-pre}
\end{align}
Here, we used $~0 < J_c/J_s < 1$ from \eq{Ic_vs_Is}.
One can show the quantity in the parenthesis is positive:
\begin{align}
   \bigg( \cdots \bigg)
  &= \int^\infty_{1} \frac{du}{u}~
      \frac{ u^{-\nu-1/2} }{ \sqrt{ (u - 1)\, W(u, 0, w) } }~
       \frac{ u - W(u, 0, w) }{W(u, 0, w)}
\nonumber \\
  &= \int^\infty_{1} \frac{du}{u}~
      \frac{ u^{-\nu-1/2} }{ \sqrt{ (u - 1)\, W(u, 0, w) } }~
       \frac{ (u - w)\, \Sigma(u)}
            {W(u, 0, w)} 
\nonumber \\
  &> 0~.
\label{eq:F_one-positivity}
\end{align}
In the final expression, we used $u \geq 1 \geq w$
and $\Sigma(u) \geq 0$ [defined in \eq{def-Sigma}].
Thus, $L_{\text{max}}$ has a local minimum at $\theta = \pi/2$.

As an example, consider the $\lambda = 1/2$ case.
This is the case for the D$p$-brane
as well as the R-charged black holes.
In this case, the resulting expressions become simple
because $W(u, 0, w) = u$ and $J_s(0, w)$
is independent of $w$.
%
%
Using \eq{L_max-pi_half}, one gets
\bea
\frac{ L_{\text{max}} }{R}
   &\sim& \frac{2\, C}{\sigma_h}~\frac{\sqrt{\pi}}{\sqrt{2\, \nu + 1}}~
    \left( \frac{2\, \nu}{2\, \nu + 1} \right)^\nu~
    \frac{ \Gamma\left( \frac{1}{2} + \nu \right) }
         { \Gamma\left( 1 + \nu \right) }
   \times \left( \frac{m \cosh^2\eta}{R^{\sigma_h}}
          \right)^{-\nu}~
\nonumber \\
  &&\times \left[~1 + \frac{\nu}{2\, (2\, \nu + 1)}
     \left( \frac{\pi}{2} - \theta \right)^2~\right]~.
\label{eq:L_max-pi_half-lambda_half}
\eea
Here, the integrals (\ref{eq:def-Js}) and (\ref{eq:def-Jc})
are evaluated as
\bea
J_s(0, w)
  &=& \int^\infty_{1} du~
      \frac{ u^{-\nu-1} }{ \sqrt{u - 1} }
  = \sqrt{\pi}~\frac{ \Gamma\left( \frac{1}{2} + \nu \right) }
                    { \Gamma\left( 1 + \nu \right) }~,
\nonumber \\
\frac{J_c(0, w)}{J_s(0, w)}
  &=& \frac{1}{2}\, \frac{ 1 + 2\, \nu }{ 1 + \nu }~.
\label{eq:Jc-lambda_half}
\eea
For the D$p$-brane, \eq{L_max-pi_half-lambda_half} indeed reduces to our previous result when $\theta=\pi/2$ (Eq.~(5.14) of Ref.~\cite{Caceres:2006ta}) using $\nu=(5-p)/(2(7-p))$ from \tabl{exponent}.

\subsection{Behavior near $\theta = 0$}\label{sec:parallel}

Next, we consider the screening length $L_s$ near $\theta=0$.
We show that the screening length at $\theta=0$
is locally a maximum generically.
This is the case for all theories which satisfy
the fall-off conditions~(\ref{eq:fall-off}).

Again, there are three constants $\hp, \hq$, and $t_c$. In this case, we first eliminate $\hp$; then eliminate the constant $\hq$ using the angle $\theta$. Finally, we vary $L$ as a function of $t_c^{-1}$ to obtain the maximum value of $L$.

First, we eliminate $\hp$ from
Eqs.~(\ref{eq:L_sin-approxII}) and (\ref{eq:L_cos-approxII})
 using Eq.~(\ref{eq:turning_pt})
\begin{align}
  & \frac{L}{R\, A}\, \sin\theta
  \simeq \hq~w^\nu~\calJ_s(\hq^2, w)~,
\label{eq:L_sin-approxIV} \\
  & \frac{L}{R\, A}\, \cos\theta
  \simeq \sqrt{ 1 - \frac{\hq^2}{1-w} }~w^\nu~
    \Big[~\calJ_s(\hq^2, w) - w~\calJ_c(\hq^2, w)~\Big]~.
\label{eq:L_cos-approxIV}
\end{align}
where $0 < w := 1/t_c \le 1$ as in the last subsection.
[The functions $\calJ_s$ and $\calJ_c$ are just $\calI_s$ and $\calI_c$, respectively, in terms of variables $(\hq^2, w)$.]
%
%
%
%
These equations can be written as
\begin{align}
  & \frac{L}{R\, A}
  \simeq \frac{w^\nu}{\cos\theta}~\sqrt{ 1 - \frac{\hq^2}{1-w} }~
    \Big[~\calJ_s(\hq^2, w) - w~\calJ_c(\hq^2, w)~\Big]~,
\label{eq:approx-Jc-zero} \\
  & \tan\theta
  \simeq \frac{\hq}{ \sqrt{ 1 - \frac{\hq^2}{1 - w} } }~
  \left( 1 - w~\frac{ \calJ_c(\hq^2, w) }{ \calJ_s(\hq^2, w) }
  \right)^{-1}~.
\label{eq:approx-tan-zero}
\end{align}

Equation~(\ref{eq:approx-tan-zero}) suggests
that $\hq = O(\theta)$,
so one can eliminate $\hq$ by $\theta$:
\begin{align}
  & \hq
  = Q(w)~\theta + O(\theta^3)~,
\label{eq:hq-approx-zero}
\end{align}
[See \eq{hq-approx-zero_app} for the definition of $Q$.]
Now, $L$ depends only on $w$ and on $\theta$
(which is fixed).
We evaluate the maximum value of $L$ as a function of $w$.

From an argument similar to \sect{perpendicular}, $L_{\text{max}}$ is given by
\bea
   \frac{ L_{\text{max}} }{R A}
  &=& G_0(w_0) + G_1(w_0)\, \theta^2 + O(\theta^4)~,
\label{eq:L-approx-zero}
\eea
where $w_0$ is determined by
\bea
  && G'_0(w_0) = 0~.
\label{eq:sol-w_star-zero0}
\eea
[See Eqs.~(\ref{eq:def-G_zero}) and (\ref{eq:def-G_one})
for the definitions of $G_0$ and $G_1$.]

Thus, what is important to us is the sign of $G_1(w_0)$.
If this is negative, then the screening length
is locally a maximum at $\theta=0$.
%
The function $G_0$ is positive-definite,
so we consider the sign of the combination $-G_1/G_0$.
Then, one can show
\begin{align}
   - \frac{G_1(w)}{G_0(w)}
  &= \frac{1}{2}\, \frac{w}{1 - w}~
    \left( 1 - \frac{\calJ_c(0, w)}{\calJ_s(0, w)} \right)^2
\nonumber \\
  &+ \frac{1}{2}\, \frac{w}{1 - w}~\frac{Q(w)}{\calJ_s(0, w)}
    \int^\infty_{1} \frac{du}{u}~\frac{ ( u - w )\, u^{-\nu-1/2} }
          { \sqrt{ (u - 1)~\calW(u, 0, w) } }\,
     \frac{1}{ \calW(u, 0, w) }
\nonumber \\
  &  - \frac{w}{2}\, \frac{\calJ_c^2(0, w)}{\calJ_s^2(0, w)}
\label{eq:G_one}
\end{align}
Because the first term on the right-hand side is positive,
\begin{align}
   - \frac{G_1(w)}{G_0(w)}
  &> \frac{w}{2\, \calJ_s(0, w)}~
    \left( \frac{Q(w)}{1 - w}~
    \int^\infty_{1} \frac{du}{u}~\frac{ ( u - w )\, u^{-\nu-1/2} }
          { \sqrt{ (u - 1)~\calW(u, 0, w) } }\,
     \frac{1}{ \calW(u, 0, w) }
        - \frac{\calJ_c^2(0, w)}{\calJ_s(0, w)} \right)
\nonumber \\
  &> \frac{w}{2\, \calJ_s(0, w)}~
    \left( \int^\infty_{1} \frac{du}{u}~
     \frac{ ( u - w )\, u^{-\nu-1/2} }
          { \sqrt{ (u - 1)~\calW(u, 0, w) } }\,
     \frac{1}{ \calW(u, 0, w) } - \calJ_c(0, w) \right)~,
\label{eq:G_one-negative-pre}
\end{align}
where we used $~0 < \calJ_c/\calJ_s < 1$ and $1 < Q(w)/(1-w)$.

One can show the quantity in the parenthesis is positive:
\begin{align}
   \bigg( \cdots \bigg)
  &= \int^\infty_{1} \frac{du}{u}~
     \frac{ u^{-\nu-1/2} }{ \sqrt{ (u - 1)~\calW(u, 0, w) } }\,
     \left( \frac{u - w}{\calW(u, 0, w)} - 1 \right)
\nonumber \\
  &= \int^\infty_{1} \frac{du}{u}~
     \frac{ u^{-\nu-1/2} }{ \sqrt{ (u - 1)~\calW(u, 0, w) } }\,
     \left( \frac{u - 1}{u^{2 \lambda} - 1} - 1 \right)
\nonumber \\
  &= \int^\infty_{1} \frac{du}{u}~
     \frac{ u^{-\nu-1/2} }{ \sqrt{ (u - 1)~\calW(u, 0, w) } }\,
      \frac{u - 1}{u^{2 \lambda} - 1}~\Sigma(u)
\nonumber \\
  & > 0~.
\label{eq:G_one-negative}
\end{align}
In the final expression, we used $u \geq 1$, $\lambda>0$, and $\Sigma(u) \geq 0$.
Thus, $L_{\text{max}}$ has a local maximum at $\theta = 0$.

\section{Discussion}\label{sec:discussion}

\subsection{Comparison with weak coupling results}\label{sec:chu-matsui}

The velocity dependence of the screening length has been analyzed by Chu and Matsui for an Abelian plasma \cite{Chu:1988wh}, so we briefly compare their results and the AdS/CFT results. Note that their results are weak coupling results (as non-Abelian plasmas) whereas the AdS/CFT computations are strong coupling results.

Chu and Matsui have computed the screened potential when an Abelian plasma is flowing relative to an electric dipole. The potential is estimated in the dipole rest frame just like our computations. They have obtained the analytic expression in the momentum space, but transforming into the coordinate space is involved, so they compute the potential numerically for $v=0, 0.5, 0.7$ and 0.9. (See Fig.~2 of Ref.~\cite{Chu:1988wh})

Their Fig.~2 shows equipotential lines of the modified Coulomb part of the gauge potential. (Note that the vector potential is also nonvanishing.) In the figure, the plasma is flowing in the $+z$-direction (horizontal direction), so the horizontal direction corresponds to $\theta=0$ and the vertical direction corresponds to $\theta=\pi/2$. 
Two features are worth mentioning: 
\begin{enumerate}
\item 
$v$-dependence: 
The contours become denser as $v\rightarrow1$. 

\item 
$\theta$-dependence: 
The contours have anisotropy in $\theta$; The lines are squashed in the horizontal direction. The anisotropy gets larger as $v\rightarrow1$.

\end{enumerate}

They did not consider the screening length itself. In fact, one would expect that their screened potential no longer takes a simple Debye potential form, so it is not clear how to define the screening length. 
In order to make a reliable comparison, we choose to discuss the $q\bar{q}$-potential instead of the screening length for the AdS/CFT case as well. For simplicity, we consider the \SAdS case for the rest of this subsection. To be self-contained, we give appropriate formulae in \appen{sads}.

\begin{figure}[tb]
\begin{center}
\includegraphics{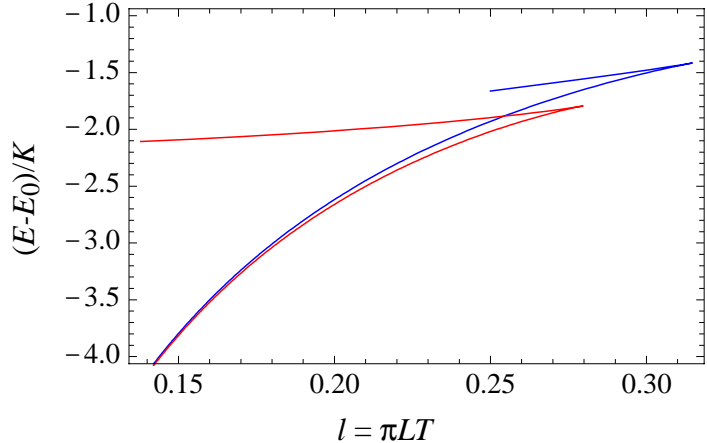}
\vskip2mm
\caption{The $q\bar{q}$-potential as a function of the dipole length $L$ ($v=0.99$). The red and blue curves represent $\theta=\pi/2$ and $\theta=0$ cases, respectively. These curves employ the scheme A for the subtraction scheme. }
\label{fig:0vs90}
\end{center}
\end{figure}

Figure~\ref{fig:0vs90} shows the $q\bar{q}$-potential for $\theta=0$ and $\theta=\pi/2$. For a given value of $L<L_s$, there exists two branches of solutions. The lower branch has a lower energy and corresponds to a larger value of the integration constant ($q>q_m$ for $\theta=\pi/2$). The upper branch has a higher energy and corresponds to a smaller value of the integration constant ($q<q_m$). It has been argued that this upper branch describes unstable solutions. 

According to \fig{0vs90}, 
\be
L(\theta=\pi/2, E)>L(\theta=0, E)
\ee
for a generic value of the potential $E$. This implies that the equipotential lines are squashed as in the weak coupling results (property~2). 
Note that $L(\theta=\pi/2, E)>L(\theta=0, E)$ but the screening length $L_s$ shows the {\it opposite} behavior, namely 
\be
L_s(\theta=\pi/2)<L_s(\theta=0)~.
\ee 
This is somewhat counter-intuitive, but the figure indeed shows that a $\theta=0$ solution exists even for a potential value where a $\theta=\pi/2$ solution does not exist. 

Next let us consider property~1 above. From this property, Chu and Matsui conclude that the screening effect becomes stronger when the pair moves. This statement itself sounds consistent with the AdS/CFT results, which predicts $L_s \propto (1-v^2)^{1/4}$, but AdS/CFT results have two complications. First, as seen in the last paragraph, the equipotential lines themselves are not a good measure of the screening length (in the strong coupling limit). Second, the energy of the dipole is given by 
\be
E= -\frac{S-S_0}{\cal T}~,
\ee
where $S$ is the action for the dipole, $S_0$ is action for a disconnected $q\bar{q}$-pair which represents the self-energy of the pair, and ${\cal T}$ is the proper time extension of the dipole. The problem is that the choice of $S_0$ is not unique as first pointed out in Ref.~\cite{Chernicoff:2006hi}. 

\begin{figure}[tb]
\begin{center}
       \includegraphics[width=7cm,height=5.0cm,clip]{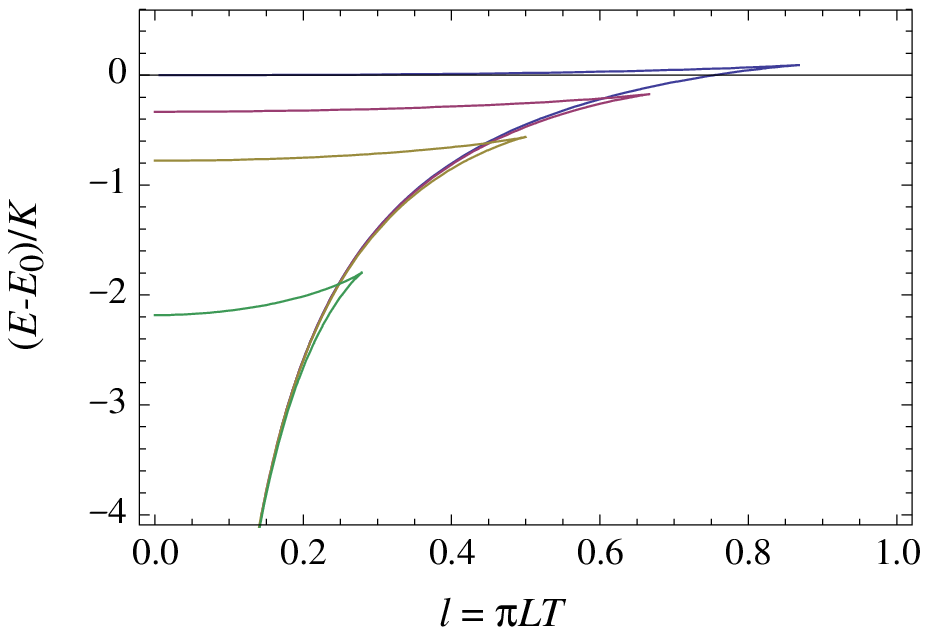}
     \hfil
       \includegraphics[width=7cm,height=5.0cm,clip]{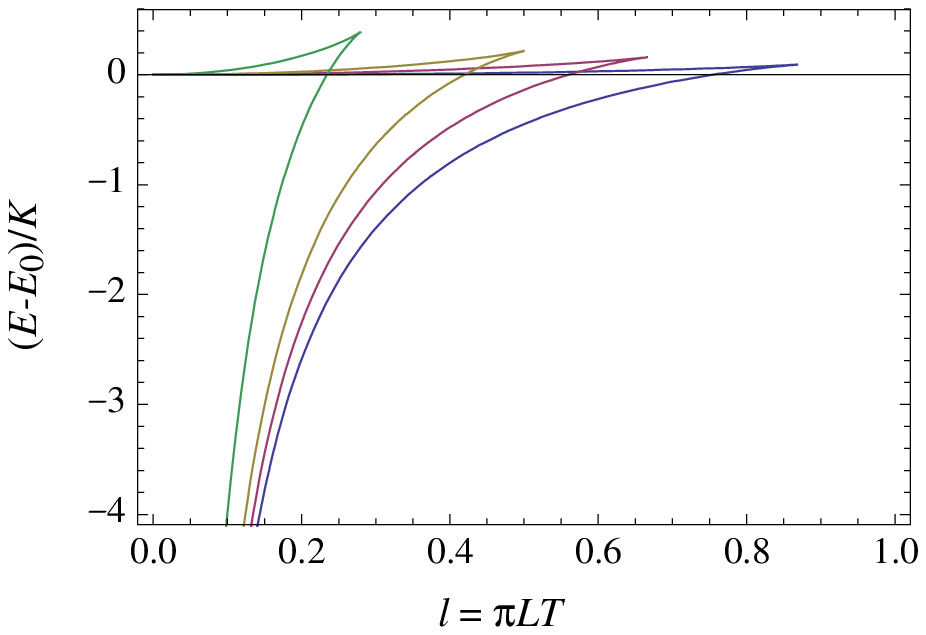}
\vskip2mm
\caption{The $q\bar{q}$-potential as a function of the dipole length $L$ for $v=0,  0.7, 0.9, 0.99$ (from right to left). The scheme A (left panel) and the scheme B (right panel) have been used for the subtraction scheme. }
\label{fig:schemes}
\end{center}
\end{figure}

This is problematic if one compares the $v$-dependence of the potential. For example, \fig{schemes} shows the potential using two different subtraction schemes $E_0$. (These schemes are reviewed in \appen{sads}.) One can see that the potential itself has little dependence on $v$ in scheme~A, whereas the $v$-dependence in scheme~B is similar to the weak coupling results. In other words, subtraction schemes in general depend on $v$, which makes the comparison unreliable. 
However, property~2 is insensitive to this problem since the subtraction schemes do not depend on $\theta$; so the comparison of $\theta$-dependence is still meaningful.

\subsection{Subleading behavior in $\eta$}\label{sec:subleading}

We have mostly considered the leading behavior $(1-v^2)^{\nu}$. For the \SAdS case, the screening length is known to have only a mild dependence on $v$ aside from the leading factor \cite{Liu:2006nn}, but it is not known if it is a generic behavior. In particular, it is interesting if the subleading terms are small even for nonconformal theories. 
 
Few theories are known when the deviation from the conformality is large, but  (the near-horizon limit of) the D$p$-brane is one such  example. For simplicity, consider the screening length for $\theta=\pi/2$. Equations~(\ref{eq:L_sin}) and (\ref{eq:L_cos}) can be integrated via hypergeometric functions. In the ultrarelativistic limit, one obtains a simple analytic expression for $L_s$ (Eq.~(5.14) of Ref.~\cite{Caceres:2006ta}), but one needs a numerical computation in general to determine $L_s$. 

One can show that the resulting $L_s$ has a mild dependence on $v$, and the leading term gives an accurate approximation.%
\footnote{We have checked this analytically using a technique similar to \sect{theta}. }
In fact, even the naive extrapolation of the ultrarelativistic limit to $v=0$ is accurate to within 20\% (\tabl{error}). The values shown in the table are the dimensionless screening length $l$ which is defined by
\bea
l &:=& \frac{4\pi}{7-p} L_s T_p~, 
\label{eq:dimless_length} \\
T_p &=& \frac{(7-p)}{4\pi}\frac{r_0^{\frac{5-p}{2}}}{R^{\frac{7-p}{2}}}~, 
\ee
where $T_p$ is the Hawking temperature for the D$p$-brane.

The D$p$-branes represent nonconformal theories, but they are often embedded as conformal theories in M-theory. The D1-brane is embedded as the M2-brane, whose near-horizon geometry is conformal ${\rm SAdS}_4$. Similarly, the D4-brane is embedded as the M5-brane, whose near-horizon geometry is conformal ${\rm SAdS}_7$. So, they are higher-dimensional conformal theories in disguise. On the contrary, the D2-brane has no such an embedding. However, the table shows that there is no significant difference for any value of $p$.

\begin{table}
\begin{center}
\begin{tabular}{|c|cccc|}
\hline
$p$		& 1	& 2	& 3	& 4 \\
\hline
$v\rightarrow0$ extrapolation of the leading term	&  0.43 &  0.54 &  0.74 &  1.2 \\
$v=0$ result	&  0.52 &  0.65 &  0.87 &  1.3 \\
$(v\rightarrow0\ {\rm extrapolation})/(v=0$ result)
		&  82\%&  84\%&  86\%&  90\% \\
\hline
\end{tabular}
\caption{The screening length for the D$p$-brane. The values represent the dimensionless screening lengths $l$ defined in \protect\eq{dimless_length}. (All values are approximate ones.) First row shows the naive extrapolation of the leading order results to $v\rightarrow0$. Second row shows the numerical results for $v=0$. Even the leading order results give good estimates for $v=0$. }
\label{table:error}
\end{center}
\end{table}

\subsection{The scaling exponent and the speed of sound}\label{sec:scaling}

In Ref.~ \cite{Caceres:2006ta}, we have proposed a relation between the scaling exponent $\nu$ and the speed of sound $c_s$. Two examples were considered: the Klebanov-Tseytlin geometry and the D$p$-brane. For both cases, the exponent and the speed of sound are related by 
\be
4\nu = 1-\frac{3}{4}(1-3c_s^2)+O\left((1-3c_s^2)\right)
\label{eq:mu_vs_cs}
\ee
when the system is nearly conformal.%
\footnote{For the D$p$-brane, we expand the D$p$-brane quantities in terms of $\deltadp:= p-3 $ and regard $\deltadp$ as a small nonconformal parameter of a 4-dimensional gauge theory. Of course, $p$ is an integer in reality.}
We do not have a proof of \eq{mu_vs_cs} for a generic system, but the following argument may be useful for the D$p$-brane.

The ultrarelativistic limit has been considered in this paper, but let us consider the opposite limit of the $v=0$ case. The exponent $\nu$ is still meaningful since the screening length takes the form
\be
L_s \propto (\epsilon_0\che)^{-\nu}~,
\ee
where $\epsilon_0$ is the energy density of (unboosted) black holes. The exponent $\nu$ is the power of the squared Lorentz factor $\che$, but one may regard the exponent $\nu$ as the power of the unboosted energy density. It is  this sense of exponent we consider below. 

Even in the $v=0$ case, the scaling argument in \sect{setup} still makes sense if the system depends only on one variable (such as \SAdSd and the D$p$-brane),%
\footnote{On the other hand, the scaling argument alone does not determine $\nu$ for R-charged black holes in the $v=0$ case, where the system depends also on a chemical potential. }
and one finds that the screening length $L_s^{(v=0)} $ is inversely proportional to temperature as in the ultrarelativistic limit. 
Then, \eq{mu_vs_cs}  is a direct consequence of
\begin{enumerate}
\item $L_s^{(v=0)} \propto T^{-1}$  
\item Standard thermodynamic relations
\end{enumerate}
as we will see below.

For the D$p$-brane, the thermodynamic variables are functions of only one variable, {\it e.g.}, the temperature $T$ since there is no chemical potential. Also, there exists a fixed mass scale $\Lambda$, and thermodynamic variables scale with appropriate powers of temperature:
\be
\epsilon_0
  = C_\epsilon~\Lambda^d~\left( \frac{T}{\Lambda} \right)^{\alpha}~, \qquad
p
  = C_p~\Lambda^d~\left( \frac{T}{\Lambda} \right)^{\beta}~, \qquad
s
  = C_s~\Lambda^{d-1}~\left( \frac{T}{\Lambda} \right)^{\gamma}~.
\label{eq:scaling_rel}
\ee
Note that thermodynamic variables have the following dimensions in $d$-dimensional spacetime: 
$[~\epsilon~] = [~p~] = {\rm M}^d$,
$[~s~] = {\rm M}^{d-1}$, and $[~T~] = {\rm M}$.
These thermodynamic variables are related to each other by the standard thermodynamic relations:%
\be
  \epsilon_0 = T\, s - p~, \qquad
  d\epsilon_0 = T\, ds ~, \qquad
  dp = s\, dT~.
\ee
Then, these equations determine exponents $\alpha$, $\beta$, and $\gamma$:
\bea
  && \alpha = \beta = \gamma + 1~,
\label{eq:exponent-relations} \\
  && C_p = \frac{C_s}{\alpha} = \frac{C_\epsilon}{\alpha - 1}~
\label{eq:coeff-relations}
\eea
Then, the sound velocity is given by
\be
  c_s^2 := \frac{\partial p}{\partial \epsilon_0} = \frac{1}{\alpha - 1}~.
\label{eq:sound_velocity}
\ee

Now, $L_s^{(v=0)} = C_L~T^{-1}$ for the D$p$-brane, so
\bea
L_s^{(v=0)} &=& C_L~T^{-1}
\nonumber \\
  &=& \frac{C_L}{\Lambda}~
   \left( \frac{\epsilon_0}{C_\epsilon\, \Lambda^d} \right)^{-1/\alpha}
\nonumber \\
  &\propto& \epsilon_0^{-1/\alpha}~.
\eea
This determines the scaling exponent $\nu$ as $\nu=1/\alpha$, and one obtains
\be
  \nu - \frac{1}{d} 
  = -\left( \frac{d-1}{d} \right)^2~
    \frac{ \frac{1}{d-1} - c_s^2 }
         { 1 - \frac{d-1}{d}\, (\frac{1}{d-1} - c_s^2) }~.
\label{eq:deviation}
\ee
This is the desired formula. If the system is nearly conformal so that $c_s^2 \sim 1/(d-1)$, expanding the above relation gives
\be
\nu - \frac{1}{d}
  \simeq \left( \frac{d-1}{d} \right)^2
         \left( \frac{1}{d-1} - c_s^2 \right)~,
\ee
or for $d=4$, 
\be
  \nu - \frac{1}{4}
  \simeq \frac{9}{16} \left( \frac{1}{3} - c_s^2 \right)~.
\label{eq:deviation_four}
\ee
This indeed agrees to \eq{mu_vs_cs}.

Actually, for the D$p$-brane, the exponent and the speed of sound are known even when the deformation is large, {\it i.e.}, $\delta_{Dp} > 1$: 
\be
\nu_{Dp}=\frac{5-p}{2(7-p)}~, \qquad
c_s^2 = \frac{5-p}{9-p}~.
\ee
These relations indeed satisfy \eq{deviation}.

\acknowledgments
We would like to thank Elena C\'aceres, Kengo Maeda, and Tetsuo Matsui for discussions. We would like to thank the Yukawa Institute for Theoretical Physics at Kyoto University and the organizers of ``String Theory and Quantum Field Theory'' and YKIS2006 "New Frontiers on QCD," where part of this work was carried out. 

\appendix

\section{Some formulae used in Sec.~3}\label{sec:formulae}

In this Appendix, we collect explicit expressions used in \sect{theta}.

\begin{itemize}

\item 
The definitions of $\calI_s$ and $\calI_s$
in \eq{L_sin-approxII} and \eq{L_cos-approxII}:
\begin{align}
  & \calI_s(\hp, \hq)
  := \int^\infty_{1} du~\frac{ u^{-\nu-1/2} }
      { \sqrt{ ( u - 1/t_c )( u^{2 \lambda} - \hp^2 ) - \hq^2\, u } }~,
\label{eq:def-calI_s} \\
  & \calI_c(\hp, \hq)
  := \int^\infty_{1} \frac{du}{u}~\frac{ u^{-\nu-1/2} }
      { \sqrt{ ( u - 1/t_c )( u^{2 \lambda} - \hp^2 ) - \hq^2\, u } }~.
\label{eq:def-calI_c}
\end{align}
The integrand of Eq.~(\ref{eq:def-calI_s}) is larger than
that of Eq.~(\ref{eq:def-calI_c}) for $u > 1$,
and both integrands are positive. So,
\be
\calI_c(\hp, \hq) < \calI_s(\hp, \hq)~.
\label{eq:Ic_vs_Is_app}
\ee

\end{itemize}

\subsection{Formulae for Sec.~3.2}

\begin{itemize}

\item 
The definition of $J_s$ and $J_c$
in \eq{L_sin-approxIII} and \eq{L_cos-approxIII}:
\begin{align}
  & J_s(\alpha, w)
  := \int^\infty_{1} du~
      \frac{ u^{-\nu-1/2} }{ \sqrt{ (u - 1) W(u, \alpha, w) } }~
  \Big( = \calI_s(\hp, \hq)  \Big)~,
\label{eq:def-Js} \\
  & J_c(\alpha, w)
  := \int^\infty_{1} \frac{du}{u}~
      \frac{ u^{-\nu-1/2} }{ \sqrt{ (u - 1) W(u, \alpha, w) } }~
  \Big( = \calI_c(\hp, \hq)  \Big)~.
\label{eq:def-Jc}
\end{align}
The function $W$ is defined by
\begin{align}
  & W(u, \alpha, w)
  := u^{2 \lambda} - ( 1 - w )~\Sigma(u) - w~\alpha~,
\label{eq:def-W} \\
  & \Sigma(u) := - u~\frac{ u^{2 \lambda-1} - 1 }{u - 1}~,
\label{eq:def-Sigma}
\end{align}
where $\alpha = \hp^2$ in the text, $\Sigma(u)$ is monotonically increasing with $u$ and $0 \le 1 - 2\, \lambda < \Sigma(u) \le 1$ for $u > 1$.

\item The definition of $P$ in \eq{hp-approx-pi_half}:
\begin{align}
  & P(w)
  := \sqrt{1-w}~
    \left( 1 - w~\frac{ J_c(0, w) }{ J_s(0, w) } \right)^{-1}~.
\label{eq:hp-approx-pi_half_app}
\end{align}

\item The definitions of $F_0$ and $F_1$ in \eq{L-approx-pi_half}:
\begin{align}
  & F_0(w)
  := \sqrt{1 - w}~w^{\nu}~J_s(0, w)~,
\label{eq:def-F_zero} \\
  & \frac{F_1(w)}{F_0(w)}
  := P^2(w)~\left[~\frac{1}{2\, P^2(w)} - \frac{1}{2}
  + \left. \frac{ \partial_{\alpha} J_s(\alpha, w)}
                {J_s(\alpha, w)} \right\vert_{\alpha=0}~\right]~.
\label{eq:def-F_one}
\end{align}

The explicit form of $L_{\text{max}}$ to order $(\pi/2-\theta)^2$
is given by
\begin{align}
   \frac{ L_{\text{max}} }{R}
  &\simeq \frac{2\, C}{\sigma_h}~\sqrt{1 - w_0}~w_0^\nu~J_s(0, w_0)
  \times \left( \frac{m \cosh^2\eta}{R^{\sigma_h}}
         \right)^{-\nu}~
\nonumber \\
  &\times 
  \left[~1 + P^2(w_0)~
    \left( \frac{1}{2\, P^2(w_0)} - \frac{1}{2}
      + \left. \frac{\partial_{\alpha} J_s(\alpha, w_0)}{J_s(\alpha, w_0)} 
         \right\vert_{\alpha=0} 
    \right) \left( \frac{\pi}{2} - \theta \right)^2~
  \right]~.
\label{eq:L_max-pi_half}
\end{align}
%

%
%

\item 
To evaluate $F_1$, one needs a derivative of $J_s$
[See \eq{def-F_one}].
The derivative can be written as
\begin{align}
  & \left. \partial_{\alpha} J_s(\alpha, w)
    \right\vert_{\alpha=0}
  = \frac{w}{2}~
      \int^\infty_{1} \frac{ du~u^{-\nu-1/2} }
                           { \sqrt{u - 1}~W^{3/2}(u, 0, w) }~.
\label{eq:relation-pi_half}
\end{align}

\end{itemize}

\subsection{Formulae for Sec.~3.3}

\begin{itemize}

\item 
The definition of $\calJ_s$ and $\calJ_c$
in \eq{L_sin-approxIV} and \eq{L_cos-approxIV}:
\begin{align}
  & \calJ_s(\alpha, w)
  := \int^\infty_{1} du~
      \frac{ u^{-\nu-1/2} }{ \sqrt{ (u - 1) \calW(u, \alpha, w) } }~
  \Big( = \calI_c(\hp, \hq)  \Big)~,
\label{eq:def-calJs} \\
  & \calJ_c(\alpha, w)
  := \int^\infty_{1} \frac{du}{u}~
      \frac{ u^{-\nu-1/2} }{ \sqrt{ (u - 1) \calW(u, \alpha, w) } }~
  \Big( = \calI_c(\hp, \hq)  \Big)~.
\label{eq:def-calJc}
\end{align}
The function $\calW$ is defined by
\begin{align}
   \calW(u, \alpha, w)
  &:= (u - w)\, \frac{ u^{2 \lambda} - 1 }{u - 1}
  + \frac{w\, \alpha}{1 - w}
\nonumber \\
  &= u^{2 \lambda} - \Sigma(u) - w\, \frac{ u^{2 \lambda} - 1 }{u - 1}
  + \frac{w\, \alpha}{1 - w}~,
\label{eq:def-calW}
\end{align}
where $\alpha = \hq^2$ in the text, and $\Sigma$ is defined in \eq{def-Sigma}.

\item The definition of $Q$ in \eq{hq-approx-zero}:
\begin{align}
  & Q(w)
  := \left( 1 - w~\frac{ \calJ_c(0, w) }{ \calJ_s(0, w) } \right)~.
\label{eq:hq-approx-zero_app}
\end{align}
Note that $1-w < Q(w) <1$ from \eq{Ic_vs_Is_app}.

\item The definitions of $G_0$ and $G_1$ in \eq{L-approx-zero}:
\begin{align}
  & G_0(w)
  := w^\nu~\Big[~\calJ_s(0, w) - w~\calJ_c(0, w)~\Big]~,
\label{eq:def-G_zero} \\
  & \frac{G_1(w)}{G_0(w)}
  := \frac{1}{2} - \frac{1}{2}\, \frac{Q^2(w)}{1-w}
      + Q^2(w)\, \left.
   \frac{ \partial_{\alpha} \big[~\calJ_s(\alpha, w)
                 - w\, \calJ_c(\alpha, w)~\big] }
        { \calJ_s(\alpha, w) - w\, \calJ_c(\alpha, w) }
     \right\vert_{\alpha=0}~.
\label{eq:def-G_one}
\end{align}

The explicit form of $L_{\text{max}}$ to order $\theta^2$
is given by
\begin{align}
   \frac{ L_{\text{max}} }{R}
  &\simeq \frac{2\, C}{\sigma_h}~w_0^\nu~
     \Big[~\calJ_s(0, w_0) - w~\calJ_c(0, w_0)~\Big]
   \times \left( \frac{m \cosh^2\eta}{R^{\sigma_h}}
          \right)^{-\nu}~
\nonumber \\
  &\times \left[~1 + \left(~\frac{1}{2}
    - \frac{1}{2}\, \frac{Q^2(w)}{1-w}
    + Q^2(w)\, \left. \frac{ \partial_{\alpha} \big[~\calJ_s(\alpha, w)
                 - w\, \calJ_c(\alpha, w)~\big] }
        { \calJ_s(\alpha, w) - w\, \calJ_c(\alpha, w) }
         \right\vert_{\alpha=0}~\right)~\theta^2~
     \right]~.
\label{eq:L_max-zero}
\end{align}

\item 
To evaluate $G_1$, one needs a derivative of $\calJ_s - w \calJ_c$. Since
\begin{align}
  & \calJ_s(\alpha, w) - w\, \calJ_c(\alpha, w)
  = \int^\infty_{1} \frac{du}{u}~\frac{ ( u - w )\, u^{-\nu-1/2} }
          { \sqrt{ (u - 1)~\calW(u, \alpha, w) } }~,
\label{eq:calJ_s-wcalJ_c}
\end{align}
the derivative can be written as
\begin{align}
  & \partial_{\alpha}~\big[~\calJ_s(\alpha, w)
     - w\, \calJ_c(\alpha, w)~\big] \big\vert_{\alpha=0}
\nonumber \\
  & \hspace*{0.5cm}
  = - \frac{1}{2}~\frac{w}{1-w}~
    \int^\infty_{1} \frac{du}{u}~\frac{ ( u - w )\, u^{-\nu-1/2} }
          { \sqrt{ (u - 1)~\calW(u, 0, w) } }\,
     \frac{1}{ \calW(u, 0, w) }~.
\label{eq:relation-zero}
\end{align}

\end{itemize}

\section{\SAdS example}\label{sec:sads}

The \SAdS black hole is given by
\be
  ds^2 = - \left(\frac{r}{R}\right)^2 
\left\{ 1-\left(\frac{r_0}{r}\right)^{4} \right\} dt^2 
+ \frac{dr^2}{\left(\frac{r}{R}\right)^2 \{ 1-(\frac{r_0}{r})^{4} \} } 
+ \left(\frac{r}{R}\right)^2 (dx_1^2+dx_2^2+dx_3^2)~,
\label{eq:SAdS5}
\ee
where $r_0$ is the horizon radius. The Hawking temperature of the black hole is given by $T=r_0/(\pi R^2)$.
For this background, Eqs.~(\ref{eq:L_sin}), (\ref{eq:L_cos}), and (\ref{eq:energy}) become 
\bea
l \sin\theta &=& 2\mathfrak{q} 
  \int^\infty_{y_c} \frac{dy}{\sqrt{\cal F}}~, \\
l \cos\theta &=&  2\mathfrak{p} 
  \int^\infty_{y_c} \frac{dy}{\sqrt{\cal F}} \frac{y^4-\che}{y^4-1}~, \\
E &=& K \int^\infty_{y_c} dy\, \frac{y^4-\che}{\sqrt{\cal F}} ~, \\
{\cal F} &:= & (y^4-\che)(y^4-1-\mathfrak{p}^2)-\mathfrak{q}^2(y^4-1)~,
\eea
where we used the dimensionless variables
\be
y := \frac{r}{r_0}~, \quad
\mathfrak{q} := \left( \frac{R}{r_0} \right)^2 q~, \quad
\mathfrak{p} := \left( \frac{R}{r_0} \right)^2 p~, \quad
l := \pi L T~. 
\label{eq:rescaled_sads}
\ee
Also, $K:= r_0/(\pi l_s^2)= \sqrt{\lambda} T$ [$\lambda=(R/l_s)^4$: 't~Hooft coupling, $l_s$: string length], and $y_c$ is the turning point given by ${\cal F}(y_c)=0$.

As usual, the energy $E$ can be made finite by subtracting the self-energy $E_0$ of a disconnected $q\bar{q}$-pair, but there are many possible choices in this case. Two schemes are often discussed in the literature, and we briefly review them here.
To compute the self-energy, choose the gauge where $\tau=t$, $\sigma=r$, and consider the configuration $x_3=x_3(\sigma)$. Then, the action $S_0$ of two disconnected quarks is given by
\be
S_0 
= - \frac{2}{2\pi l_s^2} \int d^2\sigma\, {\cal L} 
= - K{\cal T} \int dy\, \sqrt{\frac{y^4-\che}{y^4-1}
+(y^4-1) \left(\frac{r_0}{R}\right)^4 x_3'^2}~.
\ee
The conserved quantity is $q_0 := \frac{\partial {\cal L}}{\partial x_3'}$, which becomes
\be
\left(\frac{r_0}{R}\right)^4 x_3'^2 
= \frac{\mathfrak{q}_0^2}{(y^4-1)^2}\frac{y^4-\che}{y^4-\mathfrak{q}_0^2-1}~,
\label{eq:single_EOM_x3}
\ee
where $\mathfrak{q}_0$ is normalized as in \eq{rescaled_sads}. Two schemes in \sect{chu-matsui} correspond to different solutions of \eq{single_EOM_x3}:
\begin{itemize}
\item[A.]
The scheme~A uses the solution which stretches to the horizon. This condition determines $\mathfrak{q}_0$ as $\mathfrak{q}_0^2=\she$, and the energy is given by
\be
E_0 = -\frac{S_0}{\cal T} = K \int_1^{\infty} dy~.
\ee

\item[B.]
When $\mathfrak{q}_0^2<\she$, the turning point is given by $y_c^4=\che$. Then, the energy is given by 
\be
E_0 = K \int_{\sqrt{\cosh\eta}}^\infty dy\, \sqrt{\frac{y^4-\che}{y^4-\mathfrak{q}_0^2-1}}~.
\ee
In particular, the solution describes the configuration of two straight strings $x_3=({\rm constant})$ when $\mathfrak{q}_0=0$ from \eq{single_EOM_x3}. This is scheme~B.
\end{itemize}
As is clear from the expression, scheme~B is $v$-dependent. In fact, the potential in scheme~B is $v$-dependent even for small $L$. When $\cosh\eta \gg 1$ and $\mathfrak{q} \gg 1$, scheme~A gives
\be
E \sim 1 - \frac{4\pi^3}{\Gamma(1/4)^4}\frac{1}{L}~,
\ee
whereas scheme~B gives
\be
E \sim \frac{2\sqrt{2}\pi^{3/2}}{\Gamma(1/4)^2}\sqrt{\cosh\eta}
- \frac{4\pi^3}{\Gamma(1/4)^4}\frac{1}{L}~.
\ee
This motivates the authors of Ref.~\cite{Liu:2006nn} to propose that scheme~A is more natural, but the resulting potential shows a different behavior from the weak coupling results quoted in \sect{chu-matsui}.


\footnotesize

\begin{thebibliography}{99}


\bibitem{Maldacena:1997re}
  J.~M.~Maldacena,
  ``The large N limit of superconformal field theories and supergravity,''
  Adv.\ Theor.\ Math.\ Phys.\  {\bf 2} (1998) 231
  [Int.\ J.\ Theor.\ Phys.\  {\bf 38} (1999) 1113]
  [arXiv:hep-th/9711200].

\bibitem{Gubser:1998bc}
  S.~S.~Gubser, I.~R.~Klebanov and A.~M.~Polyakov,
  ``Gauge theory correlators from non-critical string theory,''
  Phys.\ Lett.\ B {\bf 428} (1998) 105
  [arXiv:hep-th/9802109].
  
\bibitem{Witten:1998qj}
E.~Witten,
``Anti-de Sitter space and holography,''
Adv.\ Theor.\ Math.\ Phys.\  {\bf 2} (1998) 253
[arXiv:hep-th/9802150].

\bibitem{Witten:1998zw}
E.~Witten,
``Anti-de Sitter space, thermal phase transition, and confinement in  gauge
theories,''
Adv.\ Theor.\ Math.\ Phys.\  {\bf 2} (1998) 505
[arXiv:hep-th/9803131].

\bibitem{Natsuume:2007qq}
  M.~Natsuume,
  ``String theory and quark-gluon plasma,''
  arXiv:hep-ph/0701201.


\bibitem{Liu:2006nn}
  H.~Liu, K.~Rajagopal and U.~A.~Wiedemann,
  ``An AdS/CFT calculation of screening in a hot wind,''
  arXiv:hep-ph/0607062.

\bibitem{Chernicoff:2006hi}
  M.~Chernicoff, J.~A.~Garcia and A.~Guijosa,
  ``The energy of a moving quark-antiquark pair in an N = 4 SYM plasma,''
  JHEP {\bf 0609} (2006) 068
  [arXiv:hep-th/0607089].

\bibitem{Peeters:2006iu}
  K.~Peeters, J.~Sonnenschein and M.~Zamaklar,
  ``Holographic melting and related properties of mesons in a quark gluon
  plasma,''
  Phys.\ Rev.\  D {\bf 74} (2006) 106008
  [arXiv:hep-th/0606195].
  

\bibitem{Caceres:2006ta}
  E.~Caceres, M.~Natsuume and T.~Okamura,
  ``Screening length in plasma winds,''
  JHEP {\bf 0610} (2006) 011
  [arXiv:hep-th/0607233].

\bibitem{Argyres:2006vs}
  P.~C.~Argyres, M.~Edalati and J.~F.~Vazquez-Poritz,
  ``No-drag string configurations for steadily moving quark-antiquark pairs in
  a thermal bath,''
  arXiv:hep-th/0608118.
  
\bibitem{Avramis:2006em}
  S.~D.~Avramis, K.~Sfetsos and D.~Zoakos,
  ``On the velocity and chemical-potential dependence of the heavy-quark
  interaction in N = 4 SYM plasmas,''
  arXiv:hep-th/0609079.

\bibitem{Talavera:2006tj}
  P.~Talavera,
  ``Drag force in a string model dual to large-N QCD,''
  JHEP {\bf 0701} (2007) 086
  [arXiv:hep-th/0610179].
 
\bibitem{Chernicoff:2006yp}
  M.~Chernicoff and A.~Guijosa,
  ``Energy loss of gluons, baryons and k-quarks in an N = 4 SYM plasma,''
  arXiv:hep-th/0611155.

\bibitem{Liu:2006he}
  H.~Liu, K.~Rajagopal and U.~A.~Wiedemann,
  ``Wilson loops in heavy ion collisions and their calculation in AdS/CFT,''
  JHEP {\bf 0703} (2007) 066
  [arXiv:hep-ph/0612168].
  

\bibitem{Kraus:1998hv}
  P.~Kraus, F.~Larsen and S.~P.~Trivedi,
  ``The Coulomb branch of gauge theory from rotating branes,''
  JHEP {\bf 9903} (1999) 003
  [arXiv:hep-th/9811120].

\bibitem{Cvetic:1999xp}
  M.~Cvetic {\it et al.},
  ``Embedding AdS black holes in ten and eleven dimensions,''
  Nucl.\ Phys.\ B {\bf 558} (1999) 96
  [arXiv:hep-th/9903214].

\bibitem{Behrndt:1998jd}
  K.~Behrndt, M.~Cvetic and W.~A.~Sabra,
  ``Non-extreme black holes of five dimensional N = 2 AdS supergravity,''
  Nucl.\ Phys.\ B {\bf 553} (1999) 317
  [arXiv:hep-th/9810227].


\bibitem{Gubser:2001ri}
  S.~S.~Gubser, C.~P.~Herzog, I.~R.~Klebanov and A.~A.~Tseytlin,
  ``Restoration of chiral symmetry: A supergravity perspective,''
  JHEP {\bf 0105} (2001) 028
  [arXiv:hep-th/0102172].

\bibitem{Buchel:2001gw}
  A.~Buchel, C.~P.~Herzog, I.~R.~Klebanov, L.~A.~Pando Zayas and A.~A.~Tseytlin,
  ``Non-extremal gravity duals for fractional D3-branes on the conifold,''
  JHEP {\bf 0104} (2001) 033
  [arXiv:hep-th/0102105].

\bibitem{Buchel:2000ch}
  A.~Buchel,
  ``Finite temperature resolution of the Klebanov-Tseytlin singularity,''
  Nucl.\ Phys.\ B {\bf 600} (2001) 219
  [arXiv:hep-th/0011146].

\bibitem{Chu:1988wh}
  M.~C.~Chu and T.~Matsui,
   ``Dynamic Debye Screening for a heavy quark-antiquark pair traversing a quark-gluon plasma,''
  Phys.\ Rev.\ D {\bf 39} (1989) 1892.
  
\end{thebibliography}
\end{document}